\documentclass[aps,pre,reprint,superscriptaddress]{revtex4-1}
\usepackage{graphicx}
\usepackage{pstricks}
\usepackage{amsmath}
\usepackage{dcolumn}

\newcommand{\beq}{\begin{equation}}
\newcommand{\eeq}{\end{equation}}

\makeatletter
\def\BState{\State\hskip-\ALG@thistlm}
\makeatother

\begin{document}
\title{An efficient GPU algorithm for tetrahedron-based Brillouin-zone integration}

\author{Daniel Guterding} 
\email{daniel.guterding@gmail.com}
\thanks{The program code is available at: https://github.com/danielguterding/cutetra}
\affiliation{Institut f\"ur Theoretische Physik, Goethe-Universit\"at Frankfurt, 
Max-von-Laue-Stra{\ss}e 1, 60438 Frankfurt am Main, Germany}

\author{Harald O. Jeschke}
\affiliation{Research Institute for 
Interdisciplinary Science, Okayama University, Okayama 700-8530, Japan}

\begin{abstract}
We report an efficient algorithm for calculating momentum-space integrals in solid state systems on modern graphics processing units (GPUs). Our algorithm is based on the tetrahedron method, which we demonstrate to be ideally suited for execution in a GPU framework. In order to achieve maximum performance, all floating point operations are executed in single precision. For benchmarking our implementation within the CUDA programming framework we calculate the orbital-resolved density of states in an iron-based superconductor. However, our algorithm is general enough for the achieved improvements to carry over to the calculation of other momentum integrals such as, e.g. susceptibilities. If our program code is integrated into an existing program for the central processing unit (CPU), i.e. when data transfer overheads exist, speedups of up to a factor $\sim130$ compared to a pure CPU implementation can be achieved, largely depending on the problem size. In case our program code is integrated into an existing GPU program, speedups over a CPU implementation of up to a factor $\sim165$ are possible, even for moderately sized workloads.
\end{abstract}


\maketitle

\section{Introduction}
When calculating the properties of periodic solid state systems, the Brillouin zone naturally appears as the smallest possible unit of momentum space that respects all spatial symmetries of the system. Many properties, like the total energy, the density of states or even magnetic susceptibilities involve integrations over the Brillouin zone. Consequently, the performance of the integration algorithm is crucial to calculating these properties efficiently.

A particularly appealing algorithm for Brillouin zone integrations employs linear approximations within tetrahedra constructed from a rectangular grid of data points in momentum space. Various tetrahedron methods have been developed to obtain the density of states~\cite{Bloechl1994, Fujiwara2002}, magnetic susceptibility~\cite{Rath1975, Heil2014}, phonon response~\cite{Savrasov1992} and electron-phonon interactions~\cite{Savrasov1996}. Recently, advances for further many-body quantities have been reported~\cite{Kawaruma2014}.

In this paper we do not aim to qualitatively improve the calculation of one of these quantities, nor do we want to add a further quantity to the list of those that can be calculated using a tetrahedron method. Instead, we present a technique to accelerate tetrahedron integration using modern graphics processing units (GPUs). We show that tetrahedron algorithms are by construction memory-bound and, therefore, well suited for GPU implementation. Indeed, we achieve speedups of up to a factor $\sim 165$.

As an example, we apply our algorithm to the problem of calculating the orbital-resolved density of states in multi-orbital systems. However, our algorithm can be applied to other quantities, such as the calculation of susceptibilities, with minimal modification. Therefore, we make our implementation in the CUDA programming framework available as an open-source code. We would like to encourage the scientific community to use our code and report any further opportunities for optimization.

We start our paper by introducing the model system that serves as an input to our algorithm. We then explain the general tetrahedron method and present our extension to integrating general momentum dependent quantities and, specifically, the orbital-resolved density of states. Subsequently, we outline our goals in designing an implementation of the tetrahedron method for GPUs and explain how we meet those goals in our specific implementation. We benchmark our implementation in two realistic scenarios and relate the benchmark results to the decisions made in our implementation. Finally, we summarize our findings.

\section{Models and Methods}
\subsection{Model Hamiltonians}
In this section we present a method for calculating the orbital-resolved density of states in real materials. A class of compounds that has a density of states with reasonably complex orbital structure, are iron-based superconductors~\cite{AndersenBoeri2011, Hosono2015a}. In these materials, the tetragonal crystal field leads to a splitting of the Fe $3d$ orbitals into four distinct groups, which determine the physics at the Fermi level. Therefore, we use these iron-based materials as an example, where the orbital-structure of the density of states is of interest.

The material of choice is FeSe due to its simple crystal structure and the availability of accurate tight-binding models. Similar to Ref.~\cite{Guterding2015a} we construct an eight-orbital tight-binding model including Fe $3d$ and Se $4p$ states. The calculated hopping parameters are denoted as $t_{ij}^{s 
p}$, where $i$ and $j$ are lattice site indices and indices $s$ and $p$ identify 
the orbitals.
\begin{equation}
\begin{split}
H_0 =& - \sum\limits_{i,j,s, p, \sigma}
t_{ij}^{s p} c^\dagger_{i s \sigma} c^{\,}_{j p \sigma}
\end{split}
\label{eq:kinetichamiltonian}
\end{equation}
This Hamiltonian can be diagonalized as a function of momentum $\vec k$ by inserting the Fourier-transform of the operators and using standard matrix diagonalization techniques. This yields band energies $E_n (\vec k)$ and matrix elements $a^m_{n} (\vec k)$ that connect orbital and band space denoted by indices $m$ and $n$ respectively. 

\begin{figure}[t]
\includegraphics[width=\linewidth]{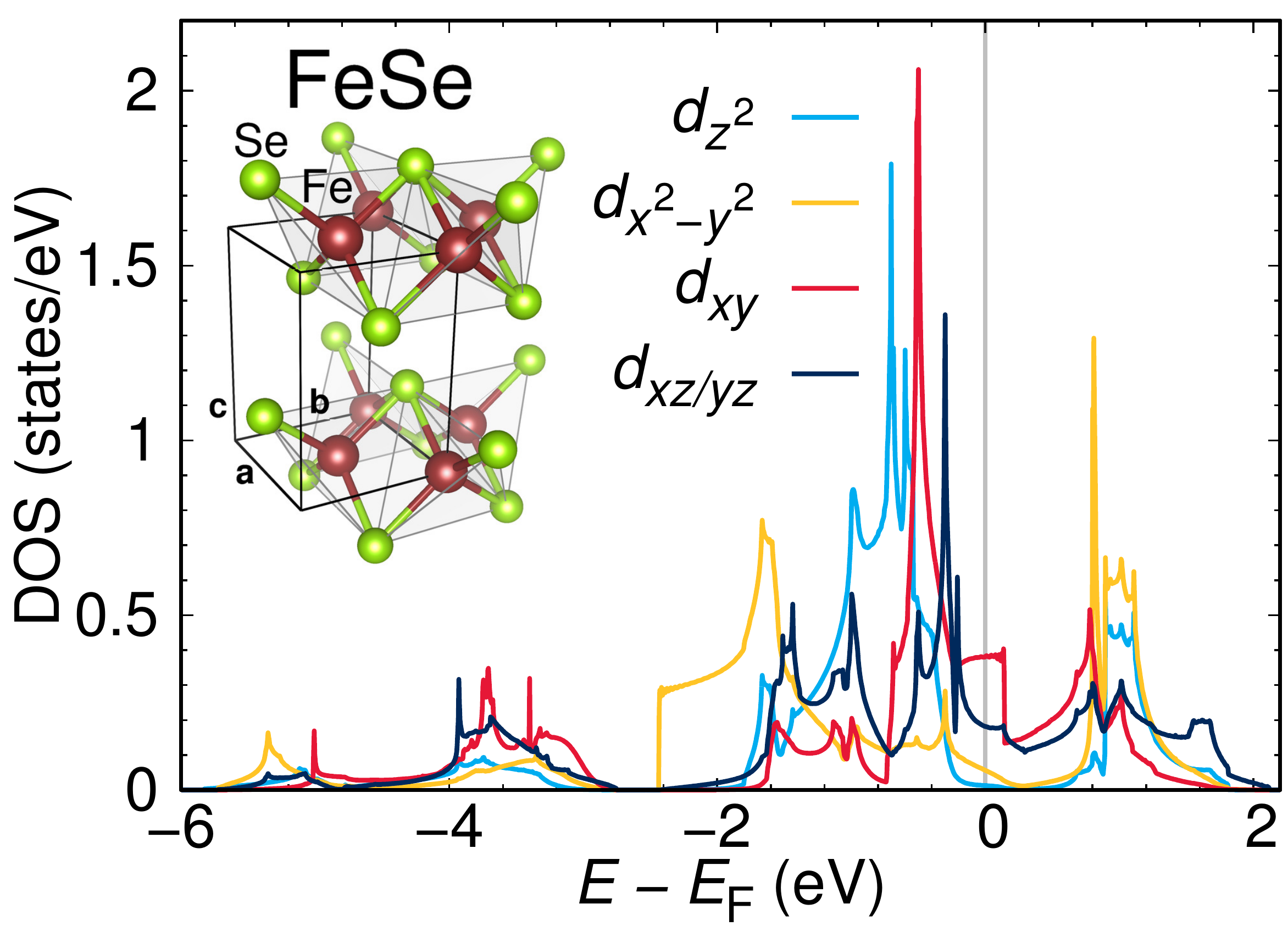}\caption{
Orbital-resolved density of states of the eight-band model for FeSe. Shown is only the density of states for the Fe $3d$ orbitals. The inset shows the crystal structure of FeSe.}
\label{fig:densityofstates}
\end{figure}

We are now interested in orbital dependent integrals over the Brillouin zone of the type
\beq
\langle X_m\rangle  = \frac{1}{V_G} \sum_n \int_{V_G}d^3k\; X_n^m(\vec{k})
f\big(E_n(\vec{k})\big)
,
\label{eq:BZintegral}
\eeq
where the occupation numbers $f(E)$ are given at $T=0$ by the Heaviside step function $\Theta(E)$, and $V_G$ is the volume of the Brillouin zone.  
The orbital-resolved particle number $n_m(E)$ would for example be obtained for the choice 
\beq
X_n^m(\vec{k})=b_{mn}(\vec k)\equiv{a^m_{n}}^\ast (\vec k)a^m_{n} (\vec k) 
.
\label{eq:matrixelements}
\eeq
Then, an energy derivative leads to the orbital-resolved density of states
\beq
\rho_m(E)= \frac{1}{V_G} \sum_n \int_{V_G} d^3k \, b_{mn}(\vec k) \delta\big(E-E_n(\vec{k})\big)
.
\label{eq:dos}
\eeq
However, we have not specified yet how to calculate the momentum integrals in practice. Our method of choice is the so-called {\it tetrahedron method}, which we explain in the following. An example for the density of states of FeSe, calculated using our method, is shown in Fig.~\ref{fig:densityofstates}.

\subsection{Generic tetrahedron method}
The general idea of the tetrahedron method~\cite{Bloechl1994} is that a quantity $\langle X \rangle$, integrated over the crystal, can be obtained as a summation over discrete $k$-points with weights $w_{n,j}$ determined by a linear interpolation over the edges of the tetrahedra.
\begin{equation}
\langle X \rangle = \sum\limits_{n} \int_{V_G} d\vec k \, X_n (\vec k) = \sum\limits_{n,j} X_n (\vec k_j) w_{n,j}
\end{equation}
Here, $j$ is the index of the $k$-point and $n$ is the band index.

The formula can be rewritten in terms of tetrahedra, labeled by index $i$, which each contain four $k$-points, i.e.
\beq
\langle X \rangle = \sum\limits_{i, n} T_{i, n}
,
\label{eq:tetmethod}
\eeq
where the tetrahedron contribution $T_{i,n}$ is given by
\beq
T_{i,n} = \sum\limits_{l=1}^{4} X_n (\vec k_l) w_{n,l}
\label{eq:tetracontrib}
\eeq
and $l$ denotes the $k$-point index within the tetrahedron.

Determination of an orbital dependent integral $\langle X_m\rangle$ now requires a generalization of this standard approach as detailed below.

\subsection{Tetrahedron integration of the orbital-resolved density of states}
In the tetrahedron method for the total density of states the quantity to integrate $X_n^m$ is set to $X_n^m = 1$ to simply count the number of states. Subsequently, the derivative with respect to the energy of interest $E$ is taken to obtain the density of states. Since $X_n^m$ is constant within the tetrahedron, the total contribution of the tetrahedron can be written down as a single expression as given in Ref.~\cite{Bloechl1994}.

In our case, we want to integrate the orbital-resolved density of states. Therefore, the quantity to integrate $X_n$ is now equal to the orbital weight $b_{n,l,m} \in [0,1]$, which does depend on momentum, even within a tetrahedron. The orbital weight $b_{n,l,m}$ is given by the square of the matrix elements $a_n^m (\vec k_l)$ as given in Eq.~\ref{eq:matrixelements}. Since the the orbital weights $b_{n,l,m}$ are not constant within a tetrahedron (they depend on momentum indexed by $l$), the contributions from the four corner points within each tetrahedron need to be summed up explicitly, contrary to the case of the total density of states. For this reason, we need to derive the integration weights $w_{n,l}$ at each of the four corner points instead of just one weight $w_n$ for the entire tetrahedron.

\begin{figure}[t]
\includegraphics[width=\linewidth]{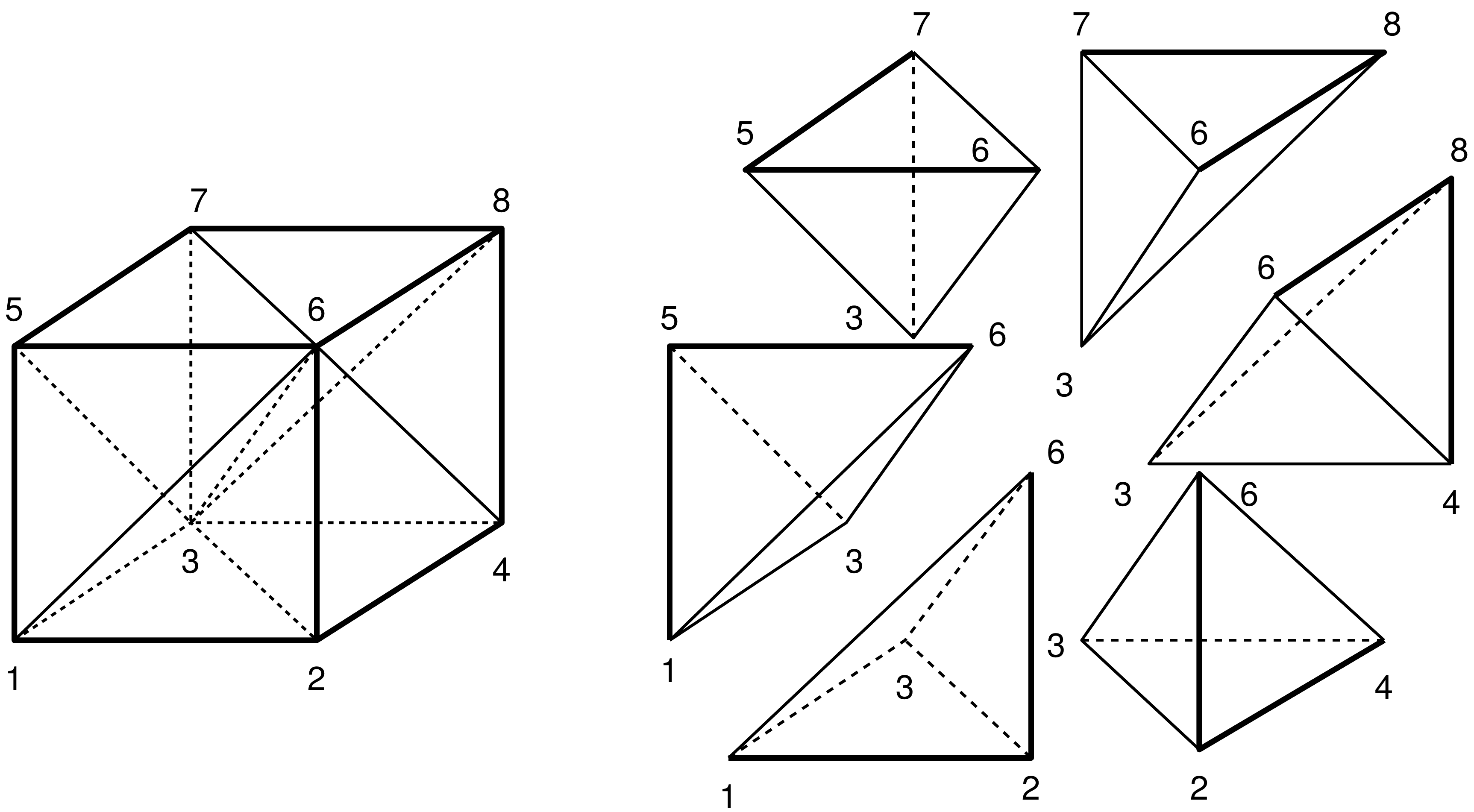}\caption{
The left-hand side of the figure shows an elemental cube of the input $k$-grid with its corner points labeled from 1 to 8. The right-hand side of the figure shows how the cube is divided into six tetrahedra. The labels correspond to the corner labels on the left-hand side.}
\label{fig:tetrahedra}
\end{figure}

For the energy dependence, note that the integration weights $w_{n,l}$ depend on the energy of interest $E$, while the orbital weights $b_{n,l,m}$ only depend on the fixed energy and momentum indices $n,l$. Consequently, taking the derivative with respect to the energy of interest $E$ in order to go from the number of states to the density of states amounts to taking the derivative of the integration weights $w_{n,l}$ only. Therefore, the tetrahedron contribution to the orbital-resolved density of states can be written as
\beq
T_{i,n,m} (E) = \sum \limits_{l = 1}^4 b_{n,l,m} \frac{\partial w_{n,l} (E)}{\partial E}
,
\label{eq:tetracontriborbitalresolved}
\eeq
where $m$ denotes the orbital. The orbital-resolved density of states $\rho_m (E)$ is then given as the sum over all tetrahedral contributions
\beq
\langle \rho_m (E) \rangle = \sum\limits_{i,n} T_{i,n,m} (E)
.
\label{eq:tetmethoorbitalresolved}
\eeq

The definitions of the tetrahedra are the same as in Ref.~\cite{Bloechl1994}. For completeness, in Fig.~\ref{fig:tetrahedra} we show how an elemental cube on the input $k$-grid is divided into six tetrahedra. The weights are calculated from linear interpolation along the edges of the tetrahedra. The band energies at constant band index $n$ at each of the four corners of a tetrahedron are sorted in ascending order and assigned labels  $E_1 \leq E_2 \leq E_3 \leq E_4$. For differences in band energies we use the abbreviation
\beq
E_{ij} = E_i - E_j
.
\eeq
The energy value of interest, i.e. the energy at which we calculate the density of states, for example, is denominated by $E$. We now list the relevant formulas for calculating the weights and their derivatives. 

In case the energy of interest is lower than or equal to the lowest energy in the tetrahedron ($E \leq E_1$), the tetrahedron does not contribute to the integral:
\beq
w_j = 0, \quad j=1,2,3,4
\eeq

\noindent In the case $E_1 < E < E_2$ we get a set of formulas:
\begin{subequations}
\allowdisplaybreaks
\begin{align}
C &= \frac{V_T}{4 V_G} \frac{(E - E_1)^3}{E_{21} E_{31} E_{41}} \\
\frac{\partial C}{\partial E} &= \frac{3 V_T}{4 V_G} \frac{(E- E_1)^2}{E_{21} E_{31} E_{41}} \\
w_1 &= C \Big[ 4 - (E - E_1) \Big( \frac{1}{E_{21}} +
 \frac{1}{E_{31}} + \frac{1}{E_{41}} \Big) \Big] \\
\frac{\partial w_1}{\partial E} &= \frac{\partial C}{\partial E} \Big[ 4 - (E - E_1) \Big( \frac{1}{E_{21}} +
 \frac{1}{E_{31}} + \frac{1}{E_{41}} \Big) \Big] \nonumber \\ 
 &- C \Big( \frac{1}{E_{21}}+ \frac{1}{E_{31}} + \frac{1}{E_{41}} \Big) \\
w_j &= C \frac{E - E_1}{E_{j1}}, \quad j=2,3,4 \\
\frac{\partial w_j}{\partial E} &= \frac{\partial C}{\partial E} \frac{E- E_1}{E_{j1}} + \frac{C}{E_{j1}} 
\end{align}
\end{subequations}

\noindent In the case $E_2 < E < E_3$ another set of formulas is used:
\begin{subequations}
\allowdisplaybreaks
\begin{align}
C_1 &= \frac{V_T}{4 V_G} \frac{(E - E_1)^2}{E_{31}E_{41}}\\
\frac{\partial C_1}{\partial E} &= \frac{V_T}{2V_G} \frac{E - E_1}{E_{31}E_{41}}\\
C_2 &= \frac{V_T}{4 V_G} \frac{(E - E_1) (E - E_2) (E_3 - E)}{E_{31} E_{32} E_{41}}\\
\frac{\partial C_2}{\partial E} &= \frac{V_T}{4V_G}\frac{1}{E_{31} E_{32} E_{41}} \Big[ (E - E_1) (E_2 - E) \\
&\quad + (E - E_1) (E_3 - E) + (E - E_2) (E_3 - E) \Big] \nonumber\\
C_3 &= \frac{V_T}{4 V_G} \frac{(E - E_2)^2 (E_4 - E)}{E_{32} E_{41} E_{42}}\\
\frac{\partial C_3}{\partial E} &= \frac{V_T}{4 V_G} \frac{E - E_2}{E_{32} E_{41} E_{42}} \Big[ 2 (E_4 - E) + (E_2 - E) \Big]\\
w_1 &= C_1 + (C_1 + C_2) \frac{E_3 - E}{E_{31}} \nonumber \\ &\quad + (C_1 + C_2 + C_3) \frac{E_4 - E}{E_{41}}\\
\frac{\partial w_1}{\partial E} &= \frac{\partial C_1}{\partial E} + \Big( \frac{\partial C_1}{\partial E} + \frac{\partial C_2}{\partial E} \Big) \frac{E_3 - E}{E_{31}} - \frac{C_1 + C_2}{E_{31}} \\& \quad + \Big( \frac{\partial C_1}{\partial E} + \frac{\partial C_2}{\partial E} + \frac{\partial C_3}{\partial E} \Big) \frac{E_4 - E}{E_{41}} \nonumber \\ & \quad - \frac{C_1 + C_2 + C_3}{E_{41}} \nonumber\\
w_2 &= C_1 + C_2 + C_3 + (C_2 + C_3) \frac{E_3 - E}{E_{32}} \\ & \quad + C_3 \frac{E_4 - E}{E_{42}} \nonumber \\
\frac{\partial w_2}{\partial E} &= \frac{\partial C_1}{\partial E} + \frac{\partial C_2}{\partial E} + \frac{\partial C_3}{\partial E} \\ \quad & + \Big( \frac{\partial C_2}{\partial E} + \frac{\partial C_3}{\partial E} \Big) \frac{E_3 - E}{E_{32}} - \frac{C_2 + C_3}{E_{32}} \nonumber \\ \quad &+ \frac{\partial C_3}{\partial E} \frac{E_4 - E}{E_{42}} - \frac{C_3}{E_{42}} \nonumber \\
w_3 &= (C_1 + C_2) \frac{E - E_1}{E_{31}} + (C_2 + C_3) \frac{E - E_2}{E_{32}}\\
\frac{\partial w_3}{\partial E} &= \Big( \frac{\partial C_1}{\partial E} + \frac{\partial C_2}{\partial E} \Big) \frac{E - E_1}{E_{31}} + \frac{C_1 + C_2}{E_{31}} \\ & \quad + \Big( \frac{\partial C_2}{\partial E} + \frac{\partial C_3}{\partial E} \Big) \frac{E - E_2}{E_{32}} + \frac{C_1 + C_3}{E_{32}} \nonumber\\
w_4 &= (C_1 + C_2 + C_3) \frac{E - E_1}{E_{41}} + C_3 \frac{E - E_2}{E_{42}} \\
\frac{\partial w_4}{\partial E} &= \Big( \frac{\partial C_1}{\partial E} + \frac{\partial C_2}{\partial E} + \frac{\partial C_3}{\partial E} \Big) \frac{E - E_1}{E_{41}} \\ & \quad + \frac{C_1 + C_2 + C_3}{E_{41}} + \frac{\partial C_3}{\partial E} \frac{E - E_2}{E_{42}} + \frac{C_3}{E_{42}} \nonumber
\end{align}
\end{subequations}

\noindent The case $E_3 < E < E_4$ is given by:
\begin{subequations}
\allowdisplaybreaks
\begin{align}
C &= \frac{V_T}{4 V_G} \frac{(E_4 - E)^3}{E_{41} E_{42} E_{43}} \\
\frac{\partial C}{\partial E} &= - \frac{3 V_T}{4 V_G} \frac{(E_4 - E)^2}{E_{41} E_{42} E_{43}} \\
w_j &= \frac{V_T}{4 V_G} - C \frac{E_4 - E}{E_{4j}}, \quad j = 1,2,3 \\
\frac{\partial w_j}{\partial E} &= - \frac{\partial C}{\partial E} \frac{E_4 - E}{E_{4j}} + \frac{C}{E_{4j}} \\
w_4 &= \frac{V_T}{4V_G} \\ &\quad - C \Big[ 4 - \Big( \frac{1}{E_{41}} + \frac{1}{E_{42}} + \frac{1}{E_{43}}\Big) (E_4 - e)  \Big] \nonumber \\
\frac{\partial w_4}{\partial E} &= -\frac{\partial C}{\partial E} \Big[ 4 - \Big( \frac{1}{E_{41}} + \frac{1}{E_{42}} + \frac{1}{E_{43}}\Big) (E_4 - e)  \Big] \\ & \quad - C \Big( \frac{1}{E_{41}} + \frac{1}{E_{42}} + \frac{1}{E_{43}} \Big) \nonumber
\end{align}
\end{subequations}

And finally, in case the energy of interest lies above the energy range of the tetrahedron ($E_4 < E$), the weights on the tetrahedron are constant:
\beq
w_j = \frac{V_T}{4V_G}, \quad \frac{\partial w_j}{\partial E} = 0, \quad j=1,2,3,4
\eeq
 
\section{GPU algorithm and implementation}
\subsection{Design goals and hardware constraints}
GPU hardware differs from usual CPUs in two important ways: (i) A GPU contains many more compute cores than a CPU, usually thousands versus a few. (ii) Memory bandwidth within the GPU is much larger than between a CPU and normal memory. 

Using the specialized hardware in the GPU comes at the expense of having to transfer data from the memory the CPU has direct access to, usually called \textit{host} memory, to the memory the GPU has direct access to, usually called \textit{device} memory. The goal of any GPU algorithm is, therefore, to use the higher memory and compute throughput of the GPU to generate large speedups that by far outweigh the overheads introduced by data transfer and synchronization between host and device code.

The usual parallelization strategy for CPU programs is to find a variable in which a formula is naturally parallel, and execute the associated program loop distributed across the available compute cores. While this strategy also works for simple algorithms when porting host code to the GPU, it is not applicable to the tetrahedron algorithm for Brillouin zone integration, as we will discuss later on. 

The reason for this is that the integration is formulated as a sum of contributions of tetrahedra, which share corners. This leads to rather complex memory access patterns, where points on the momentum grid are accessed several times. Feeding all available compute cores with enough data, therefore, becomes tricky, since the memory bandwidth even of the GPU is quickly exhausted when all compute cores work independent from each other like in the usual CPU parallelization strategy. 

\begin{figure}[t]
\includegraphics[width=\linewidth]{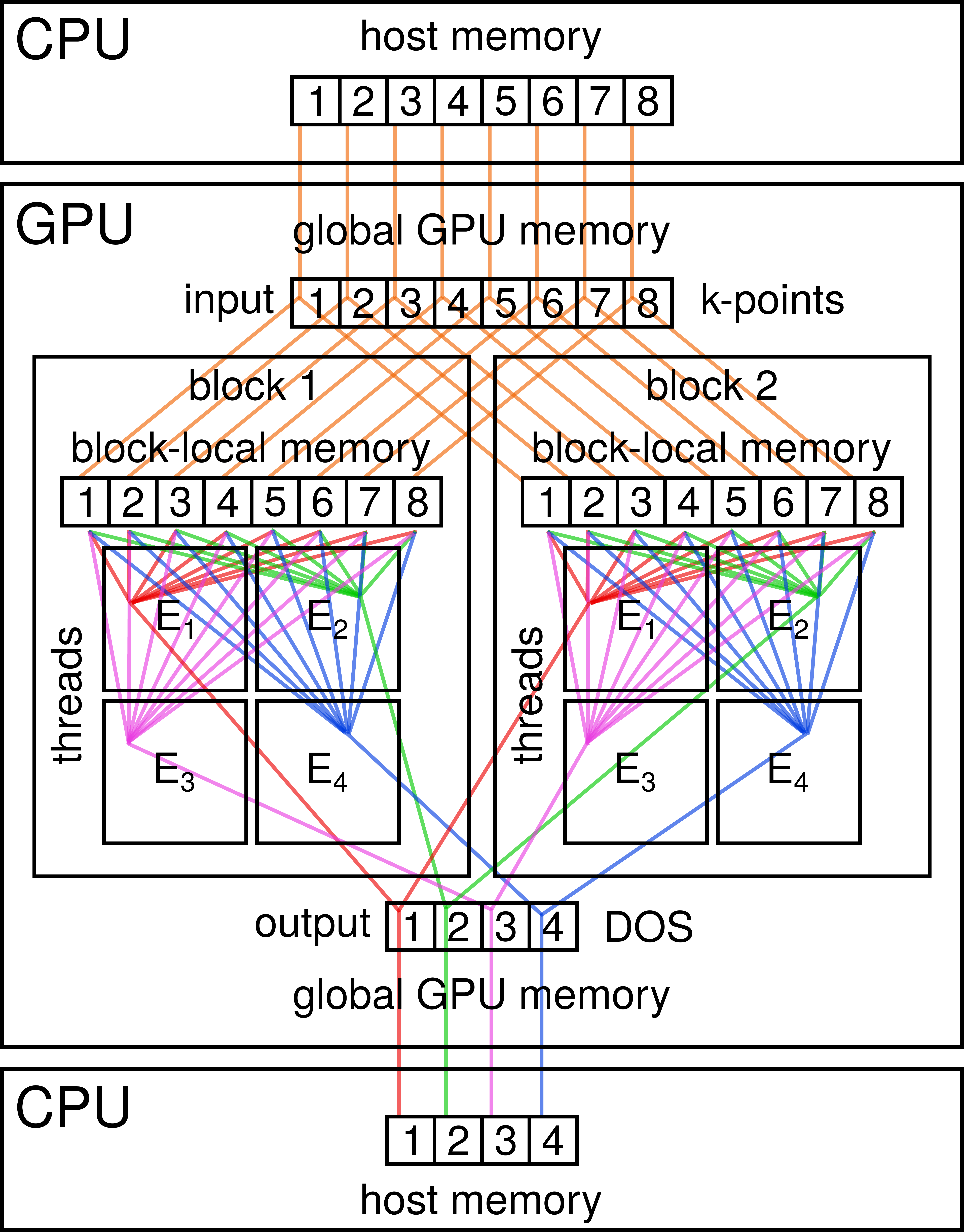}\caption{
Visualization of memory access routes in our algorithm. Cooperation of compute threads organized into blocks on common chunks of data is key to achieving high performance.}
\label{fig:memorypatterns}
\end{figure}

Fortunately, compute cores within the CUDA framework can cooperate on shared chunks of memory if they are organized into so-called \textit{blocks}. Using these blocks is essential in memory-bound problems, since block-local memory can be filled from global GPU memory once and then be accessed cheaply by all compute cores in the block. Memory access within our algorithm is visualized in Fig.~\ref{fig:memorypatterns}.

We first transfer all input data from host memory to GPU memory. Then we fetch the data that a block of threads is working on to block-local memory. Subsequently the computational kernel is executed, in which threads solely read from block-local memory and accumulate results in thread-local memory (not explicitly shown in Fig.~\ref{fig:memorypatterns}). These two are the key ingredients of the speedups we generate over a CPU implementation. After the computation is finished, results are written out to global GPU memory. An accumulation step within each block is omitted, since the results of each thread are independent in our implementation. After the results from each block have been returned, the final result accumulated over all blocks is copied to host memory.

\subsection{Algorithm outline and implementation decisions}
The algorithm for calculating the orbital-resolved density of states $\rho_m (E)$ (Eq.~\ref{eq:tetmethoorbitalresolved}) can be broken down as follows: To every elemental cube on the momentum grid we assign a CUDA block with $m$ threads. The number of external energies $E$ for which the density of states is evaluated within this block is equal to $m$. That means, all threads within the block work on the same set of data, but at different energies of interest (also referred to as frequencies). We then loop over all six tetrahedra within the elemental cube. In a further inner loop we iterate over the band index. We calculate the integration weights for this band and finally calculate the orbital-resolved density of states in a loop over orbitals.

\begin{table}[t]
\caption{Corner indices $m_j$ belonging to the tetrahedron with index $i$ within an elemental cube of the momentum grid. Compare Fig.~\ref{fig:tetrahedra}.}
\label{tab:tetradefs}
\begin{tabular}{r|rrrr}
i & $m_1$ & $m_2$ & $m_3$ & $m_4$ \\
\hline
1 & 1 & 2 & 3 & 6 \\
2 & 2 & 3 & 4 & 6 \\
3 & 1 & 3 & 5 & 6 \\
4 & 3 & 4 & 6 & 8 \\
5 & 3 & 5 & 6 & 7 \\
6 & 3 & 6 & 7 & 8 \\
\end{tabular}
\end{table}

The definition of the tetrahedra in terms of corner indices of an elemental cube in the momentum grid is given in Table~\ref{tab:tetradefs}. We put this index table into fast read-only memory, since it is a constant of our algorithm.

In the following we explain decisions we made to arrive at an efficient implementation. 
Since storing multi-dimensional arrays on GPUs is not possible without further complication, we assume band energies and orbital dependent matrix elements as a function of momentum are stored in linearized arrays. 

Therefore, on the GPU we precalculate the indices of the eight corners of every elemental cube and store them. Subsequently, for every tetrahedron we sort its four energies in ascending order and store the corresponding indices of the original array, so that we can later find the matrix elements that belong to a certain band energy. Sorting the energies within a tetrahedron is necessary in order to know which formula for the integration weights has to be used. In principle these two steps can be executed within the main computational loop, however, this would mean that index calculation and energy sorting would be performed unnecessarily by every thread in the block, although indices and energy ordering are constant per cube.

\begin{figure}[t]
\includegraphics[width=0.7\linewidth]{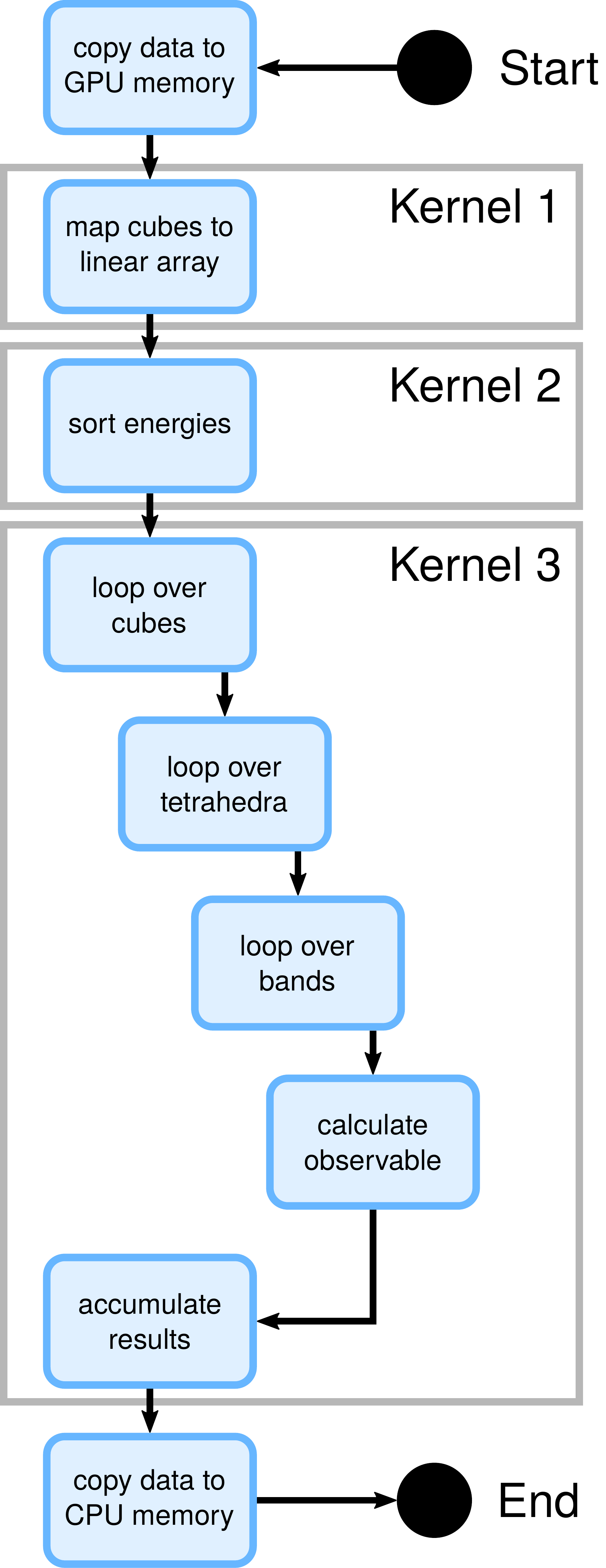}\caption{
Schematic visualization of the program flow.}
\label{fig:programflow}
\end{figure}

Instead we wrote two separate computational kernels, to be executed before the main kernel, which do index precalculation (Kernel 1) and energy sorting (Kernel 2) without wasting any computational time. Splitting a computational problem into parts that require different parallel setups is often necessary to fully optimize a GPU algorithm. The program flow is schematically visualized in Fig.~\ref{fig:programflow}.

Furthermore, Kernel 2 makes use of shared memory to store band energies for adjacent cubes on the rectangular grid. Sharing these data significantly reduces the number of load operations from global memory, since each data point is only fetched from there once, although it is used by many neighboring tetrahedra. Of course, this synergy is not perfect, since not all data can be loaded into shared memory, due to its limited size. However, the effect of sharing chunks of data is sufficient to transform Kernel 2 from being strongly memory-bound to being rather compute-bound. In our code we have used a shared memory size of two kilobytes, which is the space needed to store energies for 64 adjacent cubes at fixed band index. This quantity can be flexibly adapted to suit the needs of different hardware.

The loop over energies of interest $E$, needed for calculating the observable density of states (Kernel 3), is done in the following way: each energy is assigned a thread within the block. This way, it is guaranteed that threads within a block will never compete for write access to memory. Avoiding memory collisions is essential in avoiding wait times that slow down the computation.

However, calculations may be carried out for several cubes in parallel, which could lead to write conflicts on the central output array in global GPU memory, where the final density of states is accumulated (see Fig.~\ref{fig:memorypatterns}). Therefore, we first accumulate the results at thread-local level and then write the result out to the central result array using \textit{atomic} operations, which mediate between threads competing for write access at the same memory locations at the expense of some management overhead. 

Leaving out the thread-local accumulation of results before writing them out to global GPU memory leads to a severe decrease in performance, since the number of write collisions is roughly multiplied by the number of bands in the problem times six (the number of tetrahedra per cube), which results in significantly increased synchronization effort.

We have also implemented the possibility to let every CUDA block work on more than one elemental cube. In practice varying the number of cubes per block did not result in significantly altered performance. 

In Kernel 3, we store the eigenvector elements of the Hamiltonian at fixed band index into shared memory, since they are a shared input among all threads within the block. The purpose of using shared memory is, again, to reduce the number of slow global load operations. The size of the needed shared memory is determined by the number of orbitals in the Hamiltonian times the four corners of a tetrahedron times two, because real and imaginary parts need to be stored separately. 

For a single orbital model this yields a use of 32 bytes of shared memory per block. Therefore, the current hardware limit of 48 kilobytes per block would be exhausted at 1536 orbitals, which is well beyond the number of orbitals that researchers are typically interested in simultaneously. Of course, such large amounts of shared memory per block would also decrease performance, since less blocks could be executed in parallel. However, the performance impact would be strongly hardware specific. Therefore, we do not investigate such corner cases here.

Running Kernel 3 with shared memory allocated for 16 orbitals, instead of the five actually needed ones in our benchmark case, results in no performance loss. The reason is that Kernel 3 becomes compute rather than memory-bound once the parallel threads fetch the Hamiltonian matrix elements from shared memory. 16 orbitals correspond to the largest models of iron-based superconductors currently considered in computational solid state physics. Therefore, shared memory is not expected to significantly limit the use of our implementation in any realistic setting.

Finally, we remark that all calculations are performed in single precision, since Nvidia consumer cards do not deliver maximum performance at double precision. When specialized GPUs for scientific computations are available, one can in principle switch to double precision. However, such high precision is actually rarely needed in solid state applications.

\section{Benchmarks}
\subsection{Method}
We benchmark our GPU algorithm for the orbital-resolved density of states against an optimized equivalent CPU implementation. We use momentum grids with a different number of elemental cubes, i.e. a different resolution, to scale the workload. The smallest grid we investigate is a $10 \times 10 \times 10 = 1000$ k-point grid, the largest grid we investigate consists of $60 \times 60 \times 10 = 36000$ k-points. The grid is only refined in two dimensions, since the electronic structure of FeSe is almost two-dimensional. However, the algorithm run time is only sensitive to the total number of k-points, not their exact spatial distribution. 

We evaluate the orbital-resolved density of states on an equidistantly spaced grid of 1024 energies. The quantities we calculate are the orbital-resolved density of states for each of the five Fe $3d$ orbitals (although two of them are actually degenerate) and the total density of states.

First, the eigenenergies and eigenvectors of the tight-binding Hamiltonian are calculated on the CPU. A standard tetrahedron code for the total density of states would only require eigenenergies. However, since we are interested also in the orbital-resolved density of states, we also store the matrix elements $a^m_{n} (\vec k)$. Note that, although we here prepare the inputs for our code using a tight-binding formalism, we actually do not make any assumptions about the origin of input data. These could as well have been prepared using a density functional theory calculation.

As a reference point for measuring the speedup of the GPU algorithm, we measure the execution time of the tetrahedron integration on the CPU by taking the average execution time of five subsequent runs. To keep our point of reference as simple as possible, the CPU runs are executed on a single CPU core only. The calculation time for preparing the inputs for the tetrahedron integration, i.e. diagonalizing the Hamiltonian, is excluded here, so that the CPU execution time we measure purely represents the effort invested in the tetrahedron integration. Note that a multi-core CPU implementation could also provide large speedups compared to runs on a single CPU core. However, this parallelization method is not the focus of the present paper.

For measuring the GPU execution time we investigate two different situations: first we measure the GPU execution time including the CUDA runtime setup and the data transfer from host memory to the GPU and back from the GPU to host memory. This situation represents a worst case scenario, in which our GPU integration algorithm is integrated in an existing CPU code, where additional time has to be invested in copying data back and forth. Therefore, we refer to this situation as the \textit{worst case}. The execution time is measured in seven subsequent runs and we take the average of the five shortest run times to be the average run time. The reason behind this procedure is that consumer GPUs usually do not run at maximum clock speeds if they are not under load. Therefore, the first few runs are executed with the hardware clocked below its maximum capacity.

\begin{figure}[t]
\includegraphics[width=\linewidth]{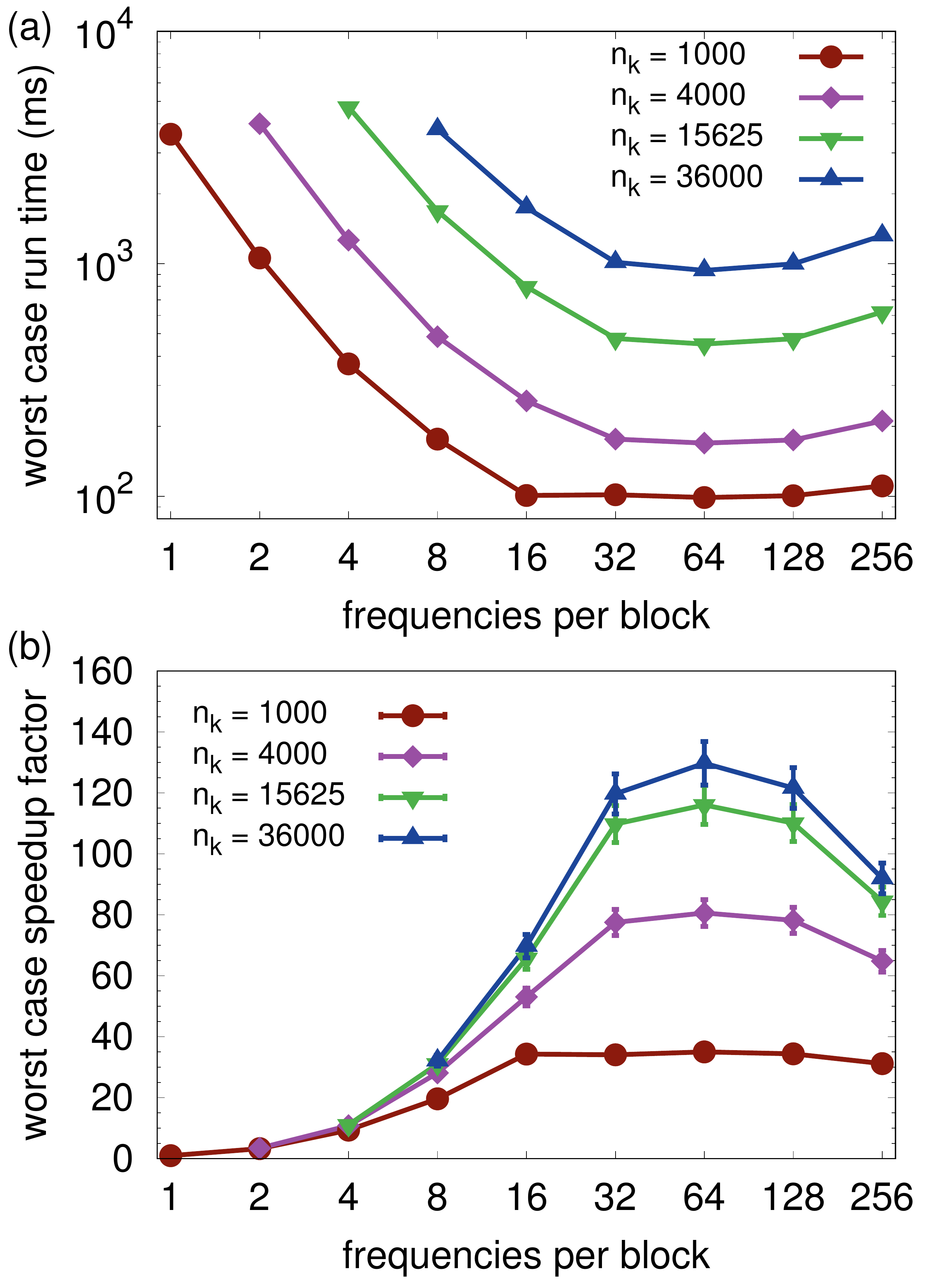}\caption{
(a) Run time in milli-seconds for the GPU algorithm in the worst case scenario for different number of momentum grid points $n_k$. (b) Speedup factor in the worst case scenario relative to the CPU implementation. Error bars correspond to an estimated maximum relative error in the speedup of $5.5~\%$.}
\label{fig:speedupworstcase}
\end{figure}

Second, we measure purely the execution time of the computational kernels on the GPU, without any CUDA runtime setup and without data transfer times to or from host memory. We refer to this situation as the \textit{best case}, which represents a best case scenario in which our algorithm is used in conjunction with an existing GPU code, where input data naturally reside in GPU memory already. Of course this implies that the worst case execution time is always larger than the best case execution time. The speedup factors are finally calculated from the average worst case/best case GPU execution time divided by the average CPU execution time. 

\begin{figure}[t]
\includegraphics[width=\linewidth]{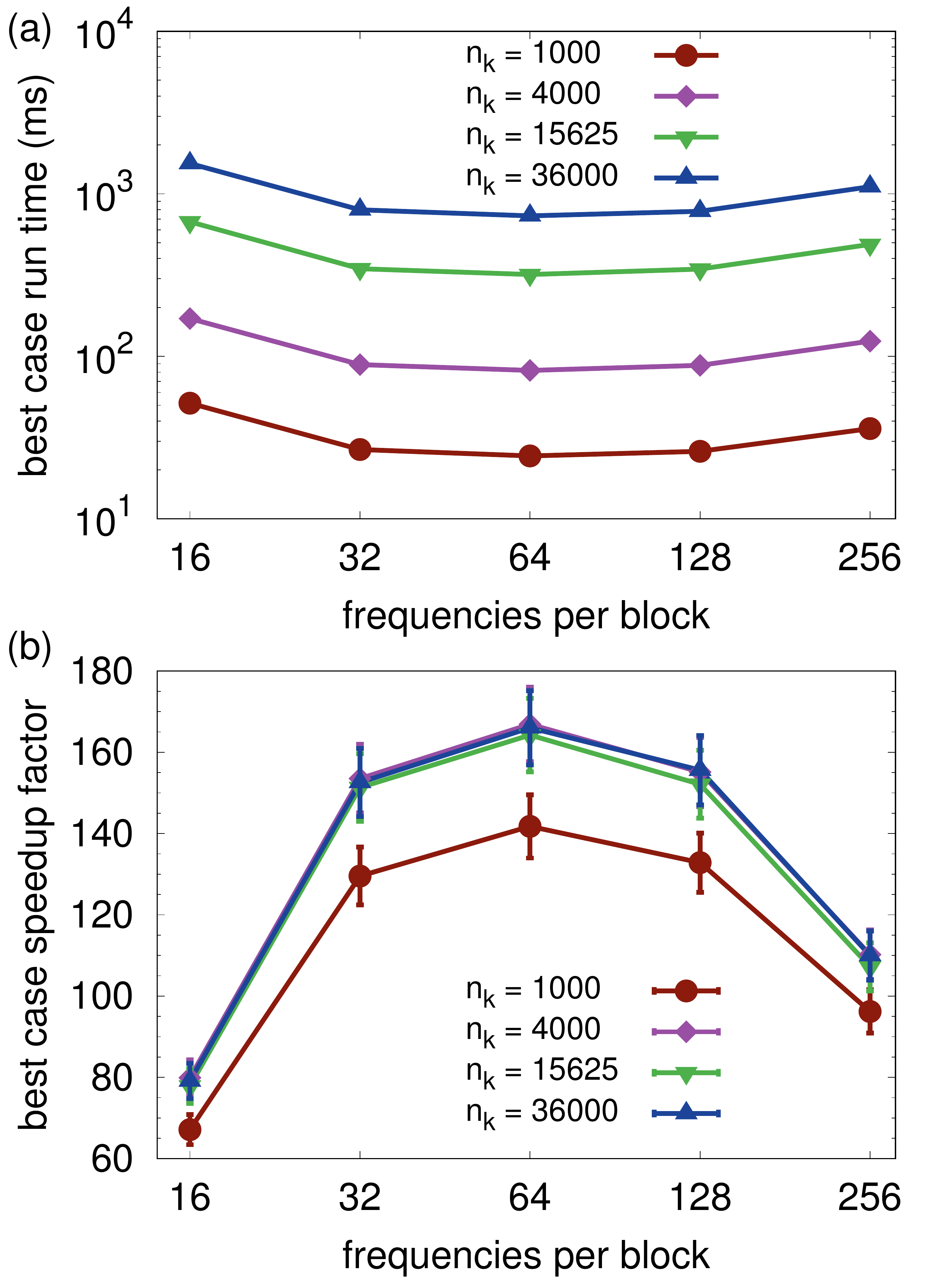}\caption{
(a) Run time in milli-seconds for the GPU algorithm in the best case scenario for different number of momentum grid points $n_k$. (b) Speedup factor in the best case scenario relative to the CPU implementation. Error bars correspond to an estimated maximum relative error in the speedup of $5.5~\%$.}
\label{fig:speedupbestcase}
\end{figure}

The CPU execution time is naturally measured in a best case situation, since our inputs were initially prepared on the CPU and are readily available. Therefore, the \textit{worst case speedup} is the lower bound for the speedup of our algorithm compared to a CPU implementation, while the \textit{best case speedup} is the upper bound.

In order to obtain an error estimate for the speedup factor, we first determine the error in measuring the program run times on CPU and GPU. We calculate the mean run time for each setting and take the maximum deviation of the measured run times within that sample from its mean as the error. The largest relative error observed across all settings is $1.45~\%$ of the run time for CPU runs and $4.04~\%$ for GPU runs. Therefore, the upper bound for the relative error in the speedup factor is $5.5~\%$. Note, however, that the actually measured error in most samples is below $1~\%$.

The hardware we tested our algorithm on employs an Intel Core i7-4790K CPU and a Nvidia GeForce 970GTX GPU. The system runs on Ubuntu Linux 16.04 with CUDA runtime version 8.0 installed.

\subsection{Results}
We vary the number of frequencies per CUDA block to find the optimal speedup for the test case we defined in the previous section. The run times and speedup factors for the worst case are shown in Fig.~\ref{fig:speedupworstcase}, while the results for the best case are shown in Fig.~\ref{fig:speedupbestcase}.

The run time increase between different work load sizes is roughly linear, as expected. We, however, are mostly interested in the dependence of execution time on the number of frequencies per block, which controls the usage of block-local memory. Fig.~\ref{fig:speedupworstcase}(b) shows that the optimal number of frequencies per block is 64, independent of the work load size. Smaller and larger numbers of frequencies per CUDA block lead to an increase in execution time. In the worst case the speedup also quite strongly depends on the size of the workload. Furthermore, for the smallest workload $n_k = 1000$ the run time is almost constant from 16 to 128 frequencies per block, which points to a significant overhead from data transfers. 

\begin{table}[t]
\caption{Run time and speedup factors as a function of the number of points on the momentum grid $n_k$. For the GPU runs we distinguish the settings worst case (GPUw) and best case (GPUb). For the GPU runs we show the results at the optimal setting of 64 frequencies per block. The speedup is calculated relative to the CPU run time.}
\label{tab:runtimes}
\begin{tabular}{r|r|r|r|r|r}
$n_k$ & CPU (ms) & GPUw (ms) & speedup & GPUb (ms) & speedup\\
\hline
1000 & 3455 & 98.7 & 35.0 & 24.4 & 141.8 \\
4000 & 13652 & 169.4 & 80.6 & 81.8 & 166.9 \\ 
15625 & 52370 & 451.4 & 116.0 & 318.9 & 164.2 \\ 
36000 & 121515 & 936.7 & 129.7 & 732.0 & 166.0
\end{tabular}
\end{table}

Therefore, we turn to the best case execution times and speedups shown in Fig.~\ref{fig:speedupbestcase}. Here, all graphs show similar behavior, independent of the workload size. For sufficiently large workload sizes, the best case speedup is practically identical in all cases. Only for the smallest input data set we find a slightly reduced speedup, because the workload is too small to put the GPU under full load.

However, also the speedup graph for the smallest workload has a clear dome shape with the maximum at 64 frequencies per CUDA block. Since the best case run time in this case is much smaller than the worst case run time, the constant behavior in Fig.~\ref{fig:speedupworstcase} simply means that the worst case run time mostly consists of data transfer time, while actual calculation time is negligible. Note, however, that even in this case our GPU algorithm performs about 35 times faster than a comparable CPU algorithm. Numerical values for run times and speedups at the optimal setting of 64 frequencies per CUDA block are also given in Table~\ref{tab:runtimes}.

The optimal number of frequencies per block, which we determined, reflects that the speedup we achieved is a compromise between fast memory access across collaborating threads on the GPU, which we increasingly exploit as the number of frequencies per block grows, and the increasing effort for synchronizing these threads as block sizes grow. The tiny speedup factors we got in case of small number of frequencies per block illustrate that it is absolutely necessary to utilize thread collaboration on GPUs to achieve significant advantages compared to CPU implementations. The usual CPU parallelization strategy, where threads work independent from each other, corresponds to the case of one frequency per block, i.e. where practically no speedup is achieved on GPU hardware.

It is possible that the optimal number of frequencies per block is different on different GPUs. However, as the speedup domes we observed are very wide, a trial-and-error search for the optimal settings on different hardware should not be difficult. Of course, GPUs with larger computational capacity may reach the optimal speedup as a function of workload size later than we did here.

\section{Summary}
We reviewed the tetrahedron method for Brillouin zone integrations and extended it to the case of the orbital-resolved density of states, supplying formulas for the momentum dependent weights within each tetrahedron.

Based on these considerations, we presented an efficient algorithm for calculating Brillouin zone integrals on modern graphics processing units and tested it on the problem of calculating the orbital-resolved density of states of an iron-based superconductor. Using the CUDA programming language we achieved an implementation that delivers large speedups of up to a factor $\sim 165$ compared to an analogous CPU implementation. We showed that optimization of memory access patterns was key to achieving these performance improvements.

Future work should concentrate on applying our algorithm to computationally expensive quantities that require Brillouin zone integrations, such as the susceptibilities employed in various many-body techniques.




\end{document}